# FIB synthesis of $Bi_2Se_3$ 1D nanowires demonstrating the co-existence of Shubnikov-de Haas oscillations and linear magnetoresistance

**Biplab Bhattacharyya**[1,2], **Alka Sharma**[1,2], **V P S Awana**[1,2], **T. D. Senguttuvan**[1,2] and **Sudhir Husale**[1,2]*

[1]Academy of Scientific and Innovative Research (AcSIR), National Physical Laboratory, Council of Scientific and Industrial Research, Dr. K. S Krishnan Road, New Delhi-110012, India.

[2] National Physical Laboratory, Council of Scientific and Industrial Research, Dr. K. S Krishnan Road, New Delhi-110012, India.

*E-mail: husalesc@nplindia.org

# Abstract

Since the discovery of topological insulators (TI), there are considerable interests in demonstrating metallic surface states (SS), their shielded robust nature to the backscattering and study their properties at nanoscale dimensions by fabricating nanodevices. Here we address an important scientific issue related to TI whether one can clearly demonstrate the robustness of topological surface states (TSS) to the presence of disorder that does not break any fundamental symmetry. The simple straightforward method of FIB milling was used to synthesize nanowires of $Bi_2Se_3$ which we believe an interesting route to test robustness of TSS and obtained results are new compared to many of the earlier papers on quantum transport in TI demonstrating robustness of metallic SS to gallium (Ga) doping. In presence of perpendicular magnetic field, we have observed the co-existence of Shubnikov-de Haas oscillations and linear magnetoresistance (LMR) which was systematically investigated at different channel lengths indicating the Dirac dispersive surface states. The transport properties and estimated physical parameters shown here demonstrate the robustness of SS to the fabrication tools triggering flexibility to explore new exotic quantum phenomena's at nanodevice level.

# 1. Introduction

Recently TIs have generated intense research in understanding the topologically non trivial quantum states and their potential applications in spintronic devices [1]. Low temperature electrical transport measurements done on nanoscale devices possess a key role in understanding many peculiar phenomena predicated with these materials but the access is limited due to isolation of surface state and their simultaneous electrical probing. One of the reasons is the presence of residual bulk carrier contributing to the surface transport [2,3]. Notably, TIs based devices demand independent metallic surface states and should have completely insulating interior bulk [4,5]. TSS have been demonstrated successfully through transport measurements such as Shubnikov–de Haas (SdH) oscillations [6], weak-anti localization (WAL) phenomena [7], high carrier mobility [8] etc. Cao et al reported the quantized hall effect and Shubnikov-de Haas oscillations in highly doped TI films ($Bi_2Se_3$) arising from the bulk of the sample but not from the surface states [9] and the related theoretical work explained the bulk contributions [10]. We show that nanostructures of TIs can be fabricated by FIB and Ga doping can be used as test to reveal the robustness of the surface states and linear MR accentuate the role of gapless linear energy dispersion. Here, by performing low temperature transport measurements, we report direct observation of quantum oscillations in $Bi_2Se_3$ nanowires which show SdH interference effects and support the quantum mechanically protected nature of topological surface states to device fabrication by FIB and gas induced metal deposition.

Our findings are relevant since it is known that topological surface states (TSS) are robust to any material deformation and nonmagnetic impurities. Previously Fukui et al. have used the FIB milling to etch $Bi_2Se_3$ films down to submicron width and studied their length dependence room temperature electrical conductivity [11]. Here we employed FIB fabricated nanowires to investigate the TSS in a TI based material which has also been deformed due to milling mechanism. Fabrication of TIs based 1D nanodevices and their transport properties shown here can be easily tuned further to make very complex nanodevices having patterned nanowires, circles, meander lines or even 3D nanostructures which are not possible with any other synthesis method. We believe that new complex nanodevices can be easily fabricated further and potentially exploited in studying the exotic electronic properties associated with TI based nanodevices.

## 2. Methods

Micromechanical exfoliation technique (scotch tape method) [12] was used to get thin flakes of $Bi_2Se_3$ on $Si/SiO_2$ chips which were already cleaned chemically (acetone, isopropanol, methanol, DI water) and with oxygen plasma (Euro plasma). Optical microscope (Olympus MX51) and FESEM was used to localize very thin flakes. Atomic force microscopy (AFM) and cross-sectional FESEM techniques have been used to calculate the thickness of the deposited flakes. Focused ion beam (FIB, Zeiss Auriga) milling (Ga ions) was used to fabricate the nanowire from $Bi_2Se_3$ thin flake. The schematic and FESEM image of milling procedure for nanowire fabrication is shown in figure 1(a). Platinum (Pt) electrodes were deposited on the nanowire using precursor based gas injection system (GIS).

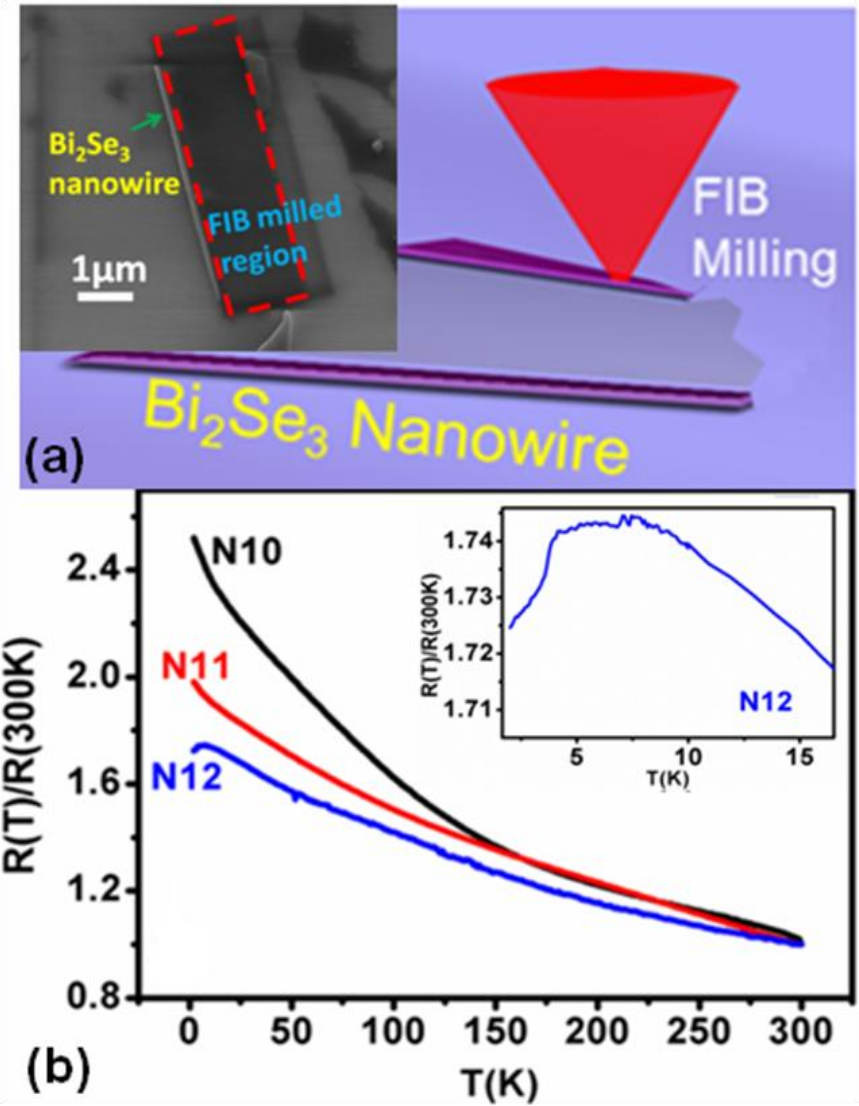

**Figure 1. Bi₂Se₃ nanowire synthesis and resistance vs. temperature (RT) characterization.** (a) Schematic of FIB milling for nanowire fabrication and inset shows the FESEM image. (b) The black, red and blue curves represent the normalized RT curves for the $Bi_2Se_3$ samples no. N10, N11 and N12 having thicknesses of 48 nm, 64 nm and 52 nm respectively. The resistance is normalized by the resistance at 300 K. Inset: The magnified RT data of sample N12 indicating a drop in resistance at T ~ 4.2 K.

## 3. Results and Discussion

We have investigated three fabricated devices (N10, 11 and 12) having similar (239 ± 9nm) nanowire widths (248, 230 and 241nm) and different channel lengths (850 nm, 150nm and 1.8 μm, distance between two voltage reading metal pads) respectively through low temperature electron transport measurements using a physical properties measurement system (PPMS by Quantum design). The normalized RT curves are shown in the figure 1(b). All three samples showed increase in resistance as a function of decrease in temperature (like semiconducting properties) but to the surprise sample N12 showed small fluctuations in resistance for the temperature range 7 to 4K and sudden drop in resistance was noticed below 4K (inset figure. 1(b)). Such observations are more common in superconducting samples or for the case where topological insulator is doped with superconductor [13]. It is not fully understood in our case for the sample N12 whether Ga milling and doping playing the role behind the observed effect because it is known that FIB deposited tungsten (combination of W, Ga and C,) nanowires shows Tc ≈ 5.2K [14] which is few orders better if compared to pure tungsten material (Tc ~11mK) [15].

To get the insights on TSS associated with fabricated nanowires, we performed MR measurements by applying an external magnetic field perpendicular to the direction of flow of current in the device and the coexistence of SdH oscillations with symmetric LMR was observed in all three samples (N10-12). The normalized MR curves measured at different base temperatures for three samples are shown in figure 2(a-b) and 3(a). A closer look at the curves for temperatures 2K, 4K and 6K, explains that resistance oscillates or fluctuates demonstrating the coexistence of SdH oscillations and LMR. The oscillations in MR curves in presence of perpendicular field has been observed earlier as Shubnikov-de Haas (SdH) oscillations in thin bismuth nanoribbons [16], bismuth selenide nanoplates [8] and bismuth telluride nanowires [17]. Further MR curves at high temperatures (≥10K) when compared to low temperatures (≤6K) indicate that oscillatory part in the MR curves increases with decrease in the temperature.

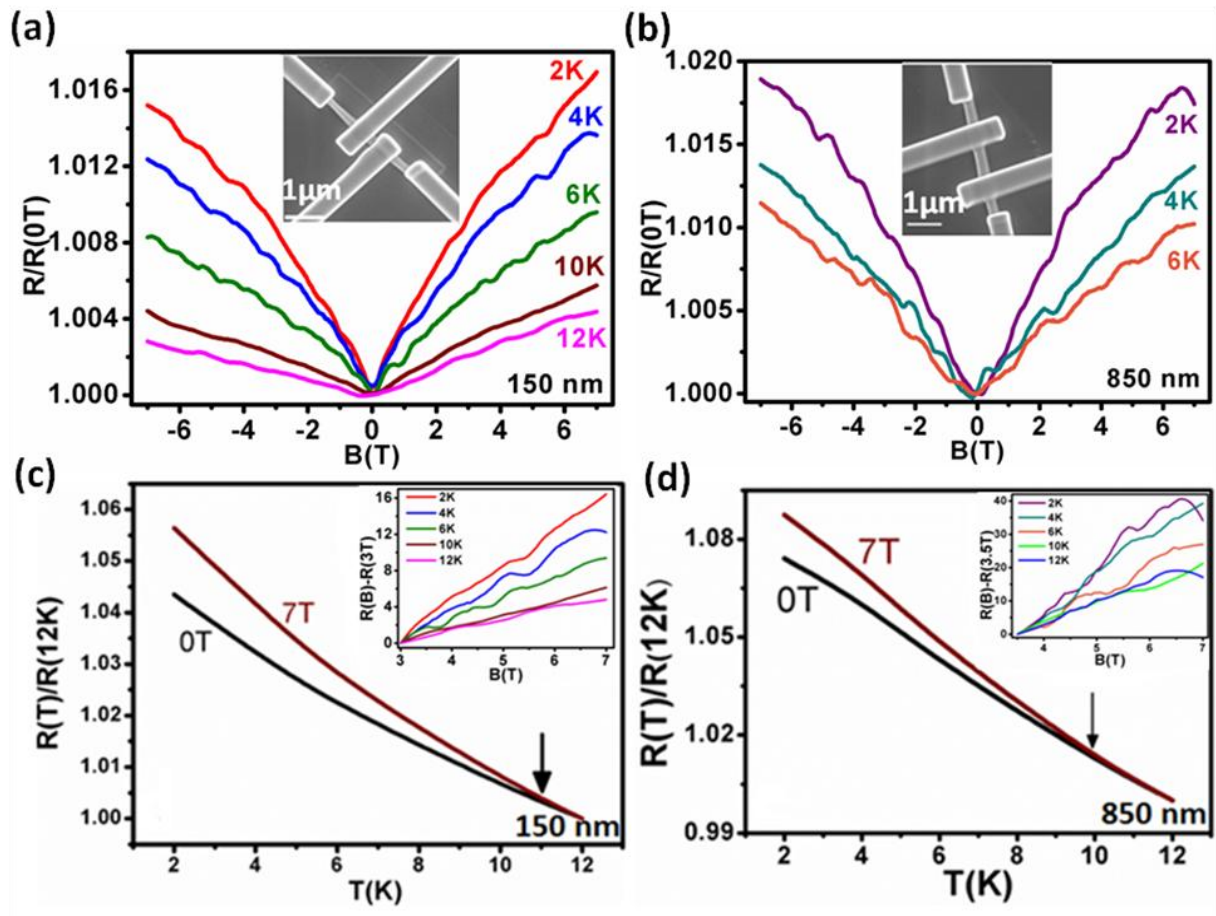

**Figure 2. Channel length dependence of LMR.** (a, b) Magnetic field dependent normalized resistance for the devices with channel lengths 150 nm and 850 nm. The corresponding FESEM images of the $Bi_2Se_3$ nanowire devices are shown in the insets. (c, d) The normalized resistances, normalized by the resistance at 12 K, vs. temperature curves extracted for CL 150 nm and 850nm as a function of B. Arrows indicate the points at which deviation in slopes started. Insets depict the isotherms of R(B)-R(3T) vs. B for 150 nm CL and R(B)-R(3.5T) vs. B for 850 nm CL.

To investigate the robust nature of TSS, here we present the channel length (CL) dependent MR analysis of the above samples (N10, N11 and N12). For the short CL (150nm, figure 2(a)), the MR at higher temperature (10K, 12K) noticeably shows linearity for the applied B field (0-7T) but MR at lower temperatures (≤ 6K) show deviation in linearity at higher fields (>3T) and this observation was consistent for CL 850 nm also (figure 2(b)). This could be due the fact that SdH oscillations originating from the surface states start to dominate LMR at high field. Strong SdH oscillations have been studied in the past at high field (B) in TI based materials and the crossover point where linearity deviates has been witnessed in other TI materials too [18,19]. R(T)/R(12K) versus T(K) curves measured for above CLs demonstrate that at high temperature (≥10K), curves measured at zero field coincides (as shown by an arrow in figure 2(c)&(d) and 3(b)) with high field B curve (7 or 6T) but at low temperature (<10K) slopes of the curves deviates. Interestingly, from the plots of R(B)-R(3T) versus B(T), taking into account the slopes of the curves, for very short CL (150 nm) we find that resistance is dependent on temperature (inset figure 2(c)) having slight dependency for higher temperature (≥10K) compared to significant changes in slopes for low temperature curves. Whereas for longer CL (1.8 μm, inset figure 3(b)) slopes do not show effective changes since all the curves overlay with each other indicating MR doesn't depend on the temperature. For the CL (850 nm, inset figure 2(d)), slopes of the curves R(B)-R(3.5T) show mixed response at different temperatures. For the temperatures ≤ 6K, the slopes of the curves upto 5T show slight changes but it starts to deviate at high fields (>5T). The MR curves investigation for larger CL (1.8 μm, figure 3(a)) at temperature > 6K show quadratic increase in the resistance as a function of B whereas MR at temperature 2K exhibit WAL effects at low fields which is shown by an arrow in inset (up). MR curves at 10K and 12K overlays completely on top of each other indicating that MR doesn't show dependency on higher temperature (≥10 K) in our measurements. Previously LMR in TIs films was found dependent on competing two

components- logarithmic phase coherence and quadratic component with a dominance of quantum phase coherence up to high temperature was observed [20]. The signature of WAL is reported coming from bulk or TSS or both is still confusing but our CLs study indicate that transport properties may become more diffusive at longer CL and specially the case of gallium doping which we have experimentally observed for the first time. From the above MR experiments performed on different CLs we observe that oscillations become more clear and distinguishable with decrease in temperature and increase in CL. The transport through longer channels manifests diffusive quantum transport in a disordered electron system and quantum interference effects are more dominant giving rise to WAL effects that forms cusp like feature. The observed WAL showing sharp cusp formation in the MR plot of fabricated nanowires was taking place at low temperature and low magnetic fields which is consistent with previous studies performed on topological insulator based material $Bi_2Te_2Se$ where they have observed high field MR is linear and nearly independent of temp over the range (T =7 to 150 K) [20]. The MR curves at 2K for above CLs have been plotted in figure 4 where the curves were shifted for clarity and following modified HLN (Hikami- Larkin-Nagaoka) fit was used.

$$\Delta G(B) = G(B) - G(0) = \alpha \frac{e^2}{\pi h}\left[\psi\left(\frac{\hbar}{4eL_\varphi^2 B}+\frac{1}{2}\right)-ln\left(\frac{\hbar}{4eL_\varphi^2 B}\right)\right]+\beta B^2 \qquad (1)$$

where, $\alpha$ = WAL coefficient. $\alpha$=0 for strong magnetic scattering, $\alpha$=1 for weak spin orbit interaction and magnetic scattering, $\alpha$=-1/2 for strong spin orbit interaction and no magnetic scattering, $e$=electronic charge=$1.6 \times 10^{-19} C$, $h$=Plank's constant=$6.626 \times 10^{-34}\ m^2\ kg/s$, $\psi$=digamma function, $\hbar$=reduced Plank's constant=$h/2\pi$, $L_\varphi$=phase coherence length (nm), $B$=magnetic field (T), $\beta$=quadratic coefficient arising from classical cyclotronic and quantum scattering terms ($\Omega^{-1}T^{-2}$). $\alpha = -0.47$ confirms the 2D nature of WAL for device N12. The fit parameter $L_\varphi(T)$ was extracted from the MR data measured for CL 1.8 μm and temperature dependent phase coherence length decay is shown in inset of figure 4.

Systematic analysis of SdH oscillations at different base temperatures was carried out for sample N12 and analysis for samples N10 and N11 has been reported in the supplementary material (figure S1 and S2). Figure 3(c) shows variation of MR as a function of inverse of magnetic field at 2K after subtraction of background. The arrows or numbers displayed on the curves were indexed to find the maxima or minima values selected to plot the Landau fan diagram as shown in the inset. The Landau-level fan diagram is used to extract the phase factor. It is known that the Landau level index (n) is related to applied magnetic field by $2\pi(n + \gamma) = S_F \frac{\hbar}{eB}$ where $\gamma$=Berry's phase, 0 or $1/2$ is the phase factor for regular electron gas or ideal topological 2D Dirac fermions, $S_F = \pi k_F^2$ = extreme cross section of the Fermi surface, $k_F$ = Fermi wave vector. The value of the finite intercept on Y axis corresponds to Berry phase ($\gamma = \frac{1}{2}$, close to half shifted behaviour) and is an important signature suggesting the existence of Dirac electrons and probability of topological surface state (TSS) transport. From the intercept at the Y axis in Landau fan diagram we have found, $\gamma$ = -0.49 for the sample N12. It is importantly to note that the observed Berry phase for samples N10 and N12 are very close to the estimated value compared in the literature that non zero Berry phase( ½ or -½) should demonstrate the transport properties of TSS. We have used the slope of the linear fit (inset figure 3(c)) to estimate the frequency ($F$) of the oscillations which is about 18T. Next the Fermi vector is estimated by using the Onsager relation (explains the dependence of oscillation frequency on the external cross section area of the Fermi surface, $A_F$)

$$\frac{1}{F} = \frac{2e}{\hbar\, k_F^2} \qquad (2)$$

and found to be 0.023 $\text{Å}^{-1}$ for the sample N12. From this we estimate the 2D and 3D surface carrier densities for the above device and which is found to be $n_{2D} = \frac{k_F^2}{4\pi} = 42.12 \times 10^{10}$ $cm^{-2}$ and $n_{3d} = \frac{k_f^3}{3\pi^2} = 411.3 \times 10^{15}$ $cm^{-3}$ respectively where $n_{2d}$ = 2D density of states, $n_{3d}$ = 3D density of states.

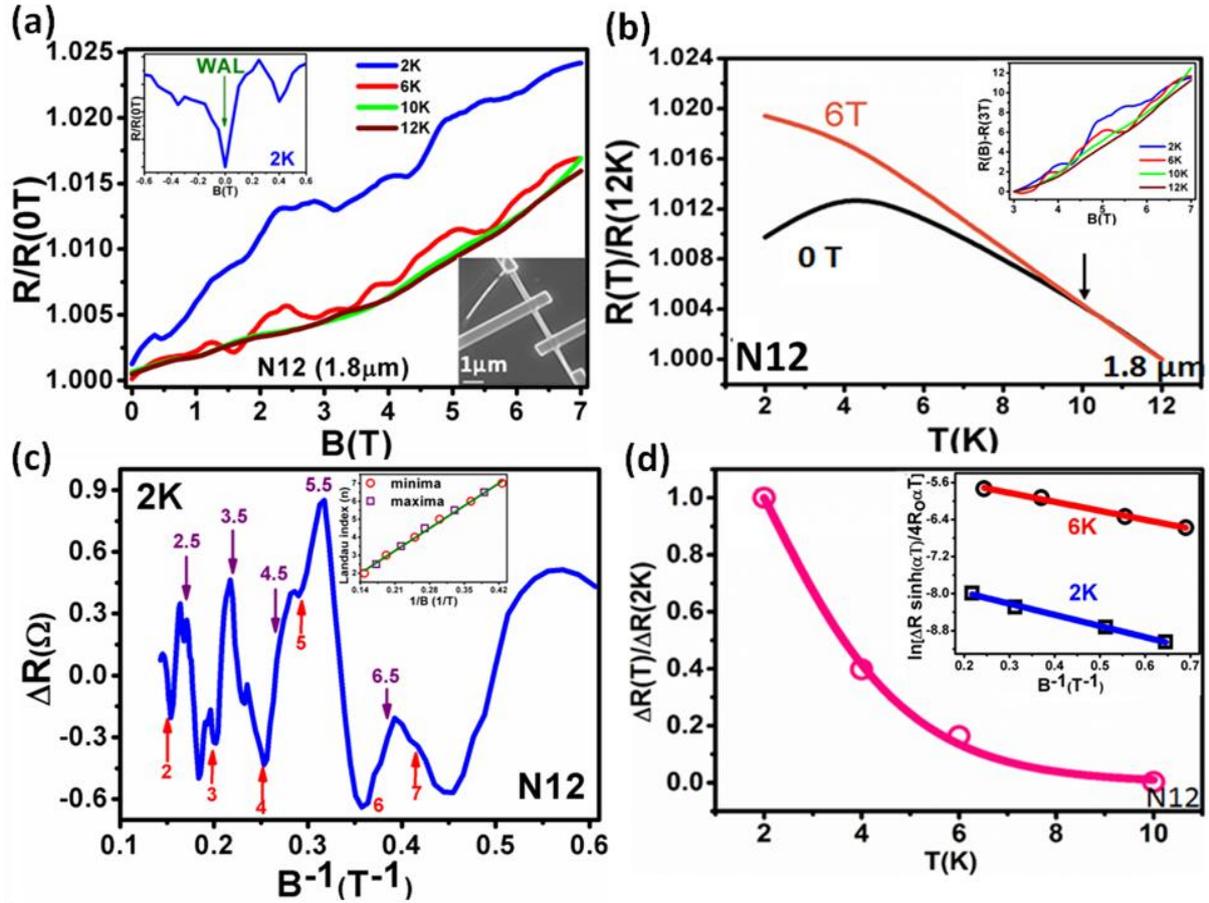

**Figure 3. Analysis of SdH oscillations in sample N12 (CL 1.8μm).** (a) Normalized MR showing WAL feature (inset) at 2K. Inset (down) shows the FESEM image of the device. (b) Normalized resistance vs. temperature curves extracted for CL 1.8μm as a function of B. Inset depict that isotherms of R(B)-R(3T) vs. B overlap for greater CL. (c) Variation in resistance (ΔR) as a function of inverse magnetic field. The arrows and numbers represent the values of maxima or minima used for plotting the Landau fan diagram as shown in the inset. (d) The temperature dependent plot of normalized resistance, normalized by the resistance measured at 2K, which represents scaled amplitude of SdH oscillations and a magnetic field 1.3 T was used to extract this data. The pink curve fit is used to estimate the cyclotron mass as explained in the text. Inset is the Dingle plot for 2K and 6K.

To get more insight into the surface state transport parameters of the above device measured under perpendicular field, the amplitude of the oscillations has been studied further to get the estimated values of cyclotron frequency $\omega_c$, cyclotron mass ($m_c$), Fermi velocity $v_f$, Fermi energy $E_f$, Dingle temperature, the transport lifetime of electrons, mean free path, mobility and metallicity parameter. For estimation of above parameters, analysis was carried out by using Lifshitz-Kosevich (LK) theory which report the field dependent oscillations in the resistance as [3,21,22]

$$\Delta R = A\, exp\left(\frac{-\pi}{\mu B}\right) cos\left[2\pi\left(\frac{F}{B} + \frac{1}{2} + \gamma\right)\right] \quad (3)$$

where $\Delta R$ =resistance fluctuation, $A$ is the temperature dependent amplitude which is proportional to $\frac{\frac{2\pi^2 k_B T}{\hbar \omega_c}}{sinh\left(\frac{2\pi^2 k_B T}{\hbar \omega_c}\right)}$, $\mu$=carrier mobility, F = slope of LL index (n) vs. $1/B$ linear fit, $\gamma$=Berry's phase, $\omega_c$=cyclotron frequency = $eB/m_c$. The cyclotron mass ($m_c$) can be obtained by using the equation

$$\frac{\Delta R(T)}{\Delta R(T_o)} = \frac{T sinh(\alpha T_o)}{T_o sinh(\alpha T)} \quad (4)$$

where $T_o$=2K. $\Delta R(T)$=resistance fluctuation at T(K), $\Delta R(T_o)$=resistance fluctuation at $T_o$(K), $T_o$=temperature (2K), $\alpha = \frac{2\pi^2 k_B}{\hbar \omega_c} = \frac{2\pi^2 k_B m_c}{\hbar e B}$, where, $k_B$ =Boltzmann constant. Figure 3(d) shows the $\frac{\Delta R(T)}{\Delta R(2K)}$ vs. T(K) graph where $\alpha$ is used as fitting parameter and B is considered as 1.3T. The corresponding best fit for $\alpha$ gives the estimated value of $m_c = 0.067\, m_e$ ($m_e$= rest or free electron mass) for the above device. Assuming the electrons are Dirac type, the Fermi velocity ($v_f$) and Fermi energy ($E_f$) are calculated for above device as $v_f = \hbar k_f/m_c = 3.99$ x $10^5$ ms$^{-1}$ and $E_f = m_c v_f^2 = 61.2$ meV respectively. The estimated Fermi velocity for sample

N12 is very consistent with Bi$_2$Te$_3$ transport studies [22] and the angle resolved photoemission spectroscopy (ARPES) studies [23]. Further the Dingle temperature estimation is done to know the life time of the surface states as $\Delta R = \frac{4*R_o*\alpha*T*e^{-\alpha T_D}}{\sinh(\alpha T)}$ where, $\Delta R$=change in resistance, $R_o$=classical resistance under zero magnetic field, $T_D$=Dingle temperature. Solving for $T_D$ we get,

$$T_D = \frac{\ln\left[\frac{\Delta R*\sinh(\alpha T)}{4R_o \alpha T}\right]}{\left(\frac{1}{B}\right)} * \left(\frac{\hbar*e}{-2\pi^2 k_B m_c}\right) = (Dingle\ plot\ slope) * \left(\frac{\hbar*e}{-2\pi^2 k_B m_c}\right) \tag{5}$$

From the slope of the inset (2K fit line) in figure 3(d), we get the $T_D$ = 2.43K. The transport lifetime of electrons or total scattering time can be estimated from the $T_D$ that is, $\tau = \frac{\hbar}{2*\pi*k_B*T_D}$ = 50.1 x 10$^{-14}$ s. Further the estimated values of mean free path, mobility and metallicity parameter are found to be $l^{SdH} = v_f \tau$ = 199.8 nm, $\mu_s^{SdH} = \frac{e\tau}{m_c}$ = 13034 cm$^2$.V$^{-1}$.s$^{-1}$ and $k_f l$ = 46.5 respectively. Note that these are estimated parameters and more experimental work needs to be performed by changing the width and thickness of the nanowire, exposing to the gallium ion doping etc. For comparison, the similar reported values of TSS have been shown in Table I in supplementary material and found that most of the parameters are in good agreement with our results.

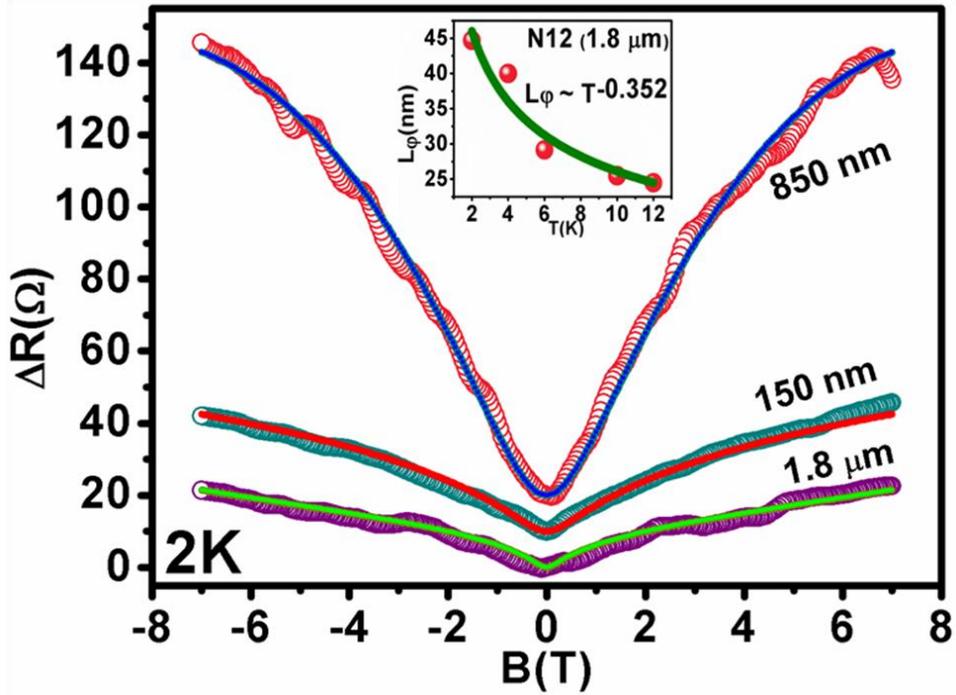

**Figure 4. Comparison of MR at 2K for the studied CLs.** The modified HLN was used to fit these curves. Inset is the phase coherence length decay plotted as a function of temperature.

One more fabricated device S1 having nanowire width ~ 640nm was studied through low temperature electron transport measurements (inset in figure 5(a)). For S1 (figure 5(a)), resistance versus temperature (RT) curve shows the metallic behaviour of the nanowire. While cooling down, resistance drops gradually with a slope of ~ 0.093 from room temperature to 36 K. A very little drop in resistance of about ~ 1 Ω is observed when lowering the temperature from 36 K to 2 K. Surprisingly, the MR measurements performed on sample S1 was found asymmetric in nature. The black curve in figure 5(b) shows a non-periodic change in resistance with B-field measured at 2K when magnetic field (B) was increased from 0 to 1.7 T as shown by the arrow. We find that fluctuations in resistance start showing some periodic behavior while further increasing B from 1.7 to 2 T. These random fluctuations in R appeared at the beginning of the measurements and were not reproducible any more. There could be multiple electron transport paths and possibilities that localization to strong disorder, accidental surface states etc. The reverse sweep of B from 2 to 0 T as shown by the red curve in figure 5(b) and its inset clearly indicates quantum oscillations in resistance with periodicity of about 0.21T (consecutive dips in the resistance). Further forward and reverse sweeps (-2T to 2T) in MR show periodic oscillations in R as shown in the inset. We find this is striking because this could happen only for electrons with extremely light mass and having large mean free path. The exact reason behind this is not known at the moment and further experimental works wait to check the reproducibility. Other studies performed on $Bi_2Se_3$ films have also reported the peculiar non symmetric LMR originating possibly from the electron surface accumulation layer of the film [24]. Switching of non-periodic to periodic oscillations behavior could be possible by the deformation mediated through ion beam milling or could be due to formation of multiband electronic structures originating from the interplay between the topology of the electronic band structure and strong spin orbit coupling in the bulk. This observed effect was not reproducible since further repetitive B sweeps (-2 to 2T) didn't show any aperiodicity and this gives a hint that 2D metallic surface states conduction is dominating factor in fabricated nanodevices (S1). Previously a peculiar non-symmerty in MR was observed in high-index (221) $Bi_2Se_3$ films and origin of disorder was predicated possibly arising from the electron surface accumulation layer [24].

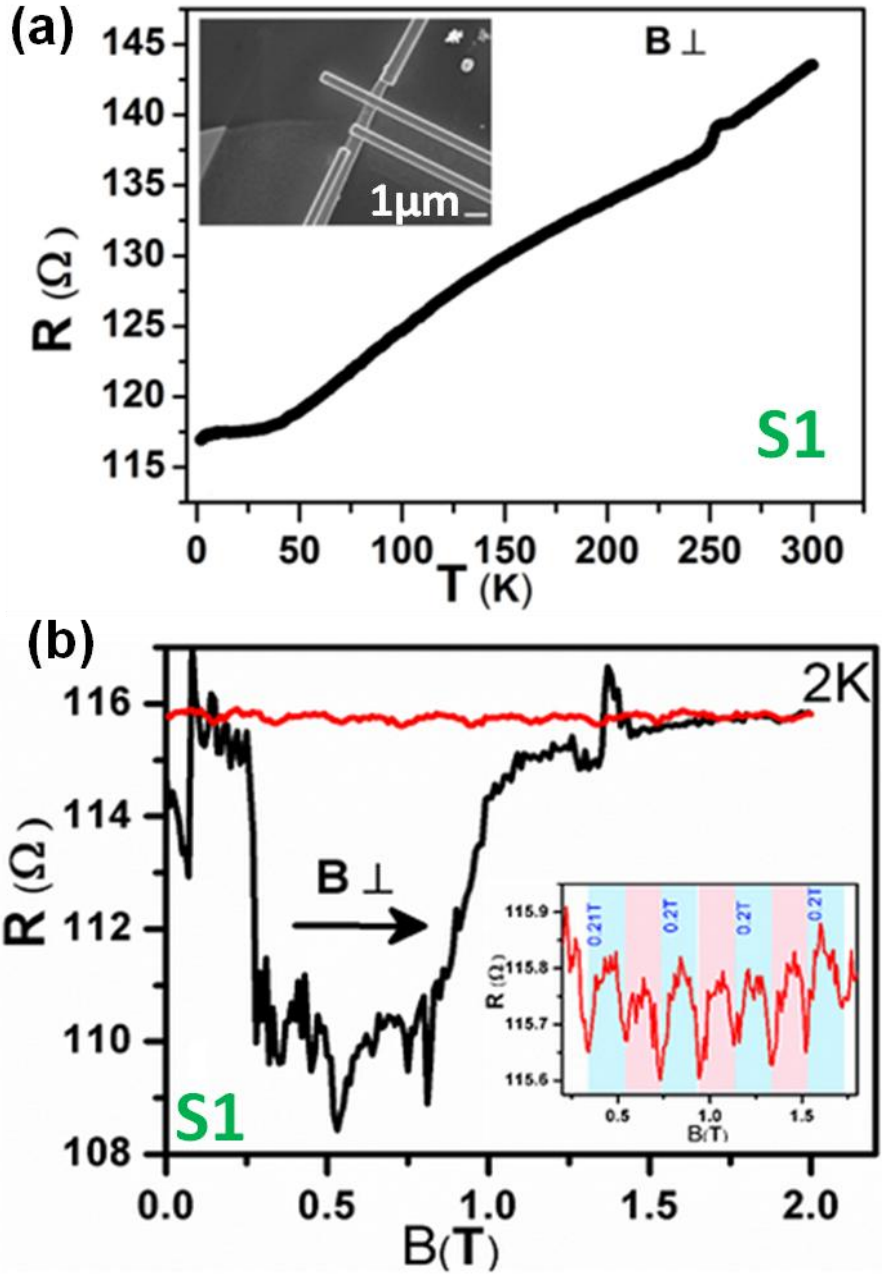

**Figure 5. Sample S1.** (a) RT curve shows metallic behavior, inset is device S1. (b) SdH oscillations in FIB fabricated device S1, a non periodic to periodic oscillations in presence of perpendicular magnetic field at the base temperature 2K. The arrow indicates sweep direction of B. Up sweep (0 to 2T, the black curve) and down sweep (2-0T, the red curve). Inset shows the periodic oscillations in MR observed for next up or down sweeps in B.

The devices used here were fabricated by FIB milling using Ga to make the nanowire from the flake which may implant some Ga impurities in the material which is unavoidable. Even though very precise slices of flakes can be achieved with FIB operating with low milling current (< 50pA) but there is a chance that milled out material might get slightly deformed. Additionally to make electrical pads of Pt for low temperature electrical transport studies, the GIS were employed which uses gaseous phase of molecules that may deposit some contamination [25]. It is very difficult to determine the amount of contamination at nanostructure level but transport signatures through these devices show robustness of TSS and the related properties do not get altered by device fabrication technique. From the experimental analysis, we conclude that devices investigated here demonstrate the existence of TSS in TIs based FIB fabricated nanoscale devices. Previously in TI materials like $Bi_2Se_3$, WAL effects observed in MR was correlated to 2D whereas the work on In-doped SnTe nanoplates revealed the dependence of surface state of topological crystalline insulator on the level of doping and the observed 2D linear MR at high magnetic field was attributed to Dirac dispersive states [13]. Further they reported that the observed WAL in $In_xSn_{1-x}Te$ nanoplates is a bulk effect. Our fabricated nanowire sample (figure 3(a), N12) does show a sharp cusp but only at low temperature and low magnetic field. At large magnetic fields and temperatures >4K WAL effect was not observed. Note that in agreement with our LMR results, recently such WAL effects was observed at low magnetic fields in In-doped SnTe and their angle dependence (0 to 90°) MR measurements revealed that WAL feature was overlapping perfectly for all angles of the B and thus it was mentioned that WAL feature is a 3D bulk effect. Further in agreement with their results we have also observed that above 10K WAL feature disappears (figure 3(a)). The observed LMR in short CL (figure 2(a)) could be due to the Dirac-dispersive nature of surface states which is the property of TIs. The possibility of LMR was earlier proposed by Abrikosov in gapless linear dispersive energy spectrum under the condition of first landau level is filled, indicating the possibility of non-saturating linear MR [26], whereas Wang and Lei theory suggested the presence of gapless linear spectrum overlapping with Landau level [27]. Further, we have observed the slope of the LMR (inset figure 3(b), sample N12) is temperature independent which is in agreement with the experimental findings of temperature independent LMR observed in In doped SnTe.

## 4. Conclusion

We show a simple method of TIs based nanowire fabrication from high quality micromechanically exfoliated flakes. The low temperature transport studies through RT measurements showed semiconducting (N10-12) behaviour. From the Landau plot, we have observed SdH oscillations close to ½ shifted, indicating the 2D metallic surface states. The parameters estimated from the MR oscillations are in the order similar to those published in the literature. Finally, we have found that quantum oscillations in FIB fabricated nanowire survive even after exposed to a rigorous milling process using Ga ion and to gas phase metal deposition process indicating the signatures of topological states and their robustness. We emphasize here that it is very likely for many topological insulator based devices can be fabricated using our approach to study the surface transport properties at 1D/2D level and can be exploited for investigating interesting fundamental problems or exotic physics like magnetic monopole, Majorana fermions etc. in near future .

# Acknowledgements

S.H. and A.S. acknowledge the financial support of CSIR's Network project "Aquarius". B.B. acknowledges the support of CSIR-Junior Research Fellowship. We thank Dr. Sangeeta Sahoo for useful discussions and critical reading of the manuscript. We thank Dr. Ranjana Mehrotra and Dr. V. N. Ojha for their guidance and support.

**Figures:**

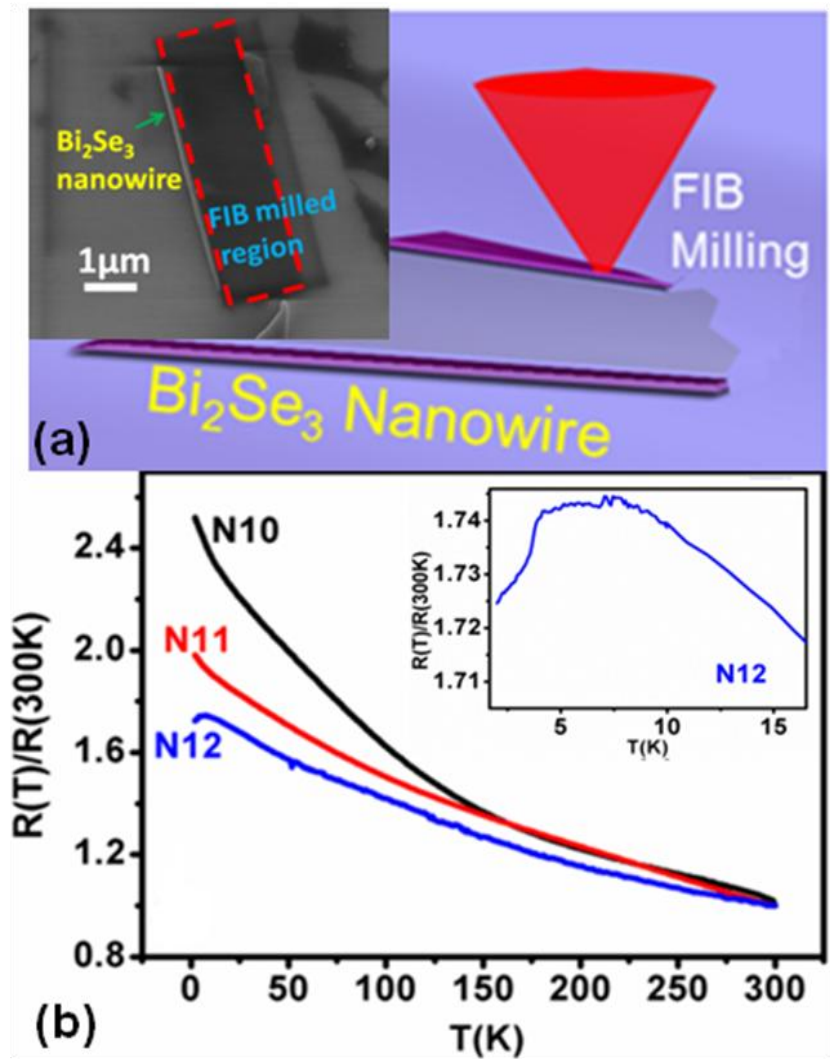

**Figure 1. $Bi_2Se_3$ nanowire synthesis and resistance vs. temperature (RT) characterization.** (a) Schematic of FIB milling for nanowire fabrication and inset shows the FESEM image. (b) The black, red and blue curves represent the normalized RT curves for the $Bi_2Se_3$ samples no. N10, N11 and N12 having thicknesses of 48 nm, 64 nm and 52 nm respectively. The resistance is normalized by the resistance at 300 K. Inset: The magnified RT data of sample N12 indicating a drop in resistance at T ~ 4.2 K.

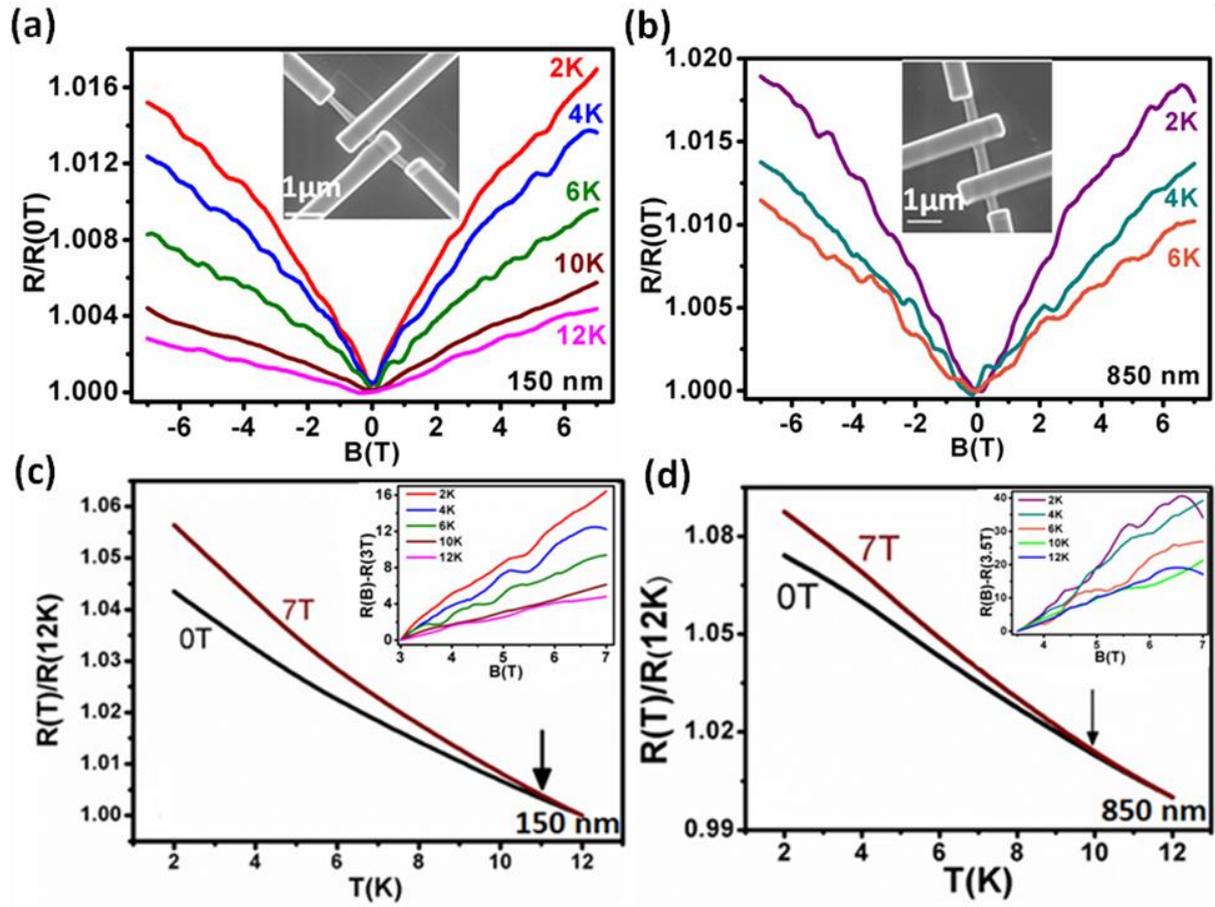

**Figure 2. Channel length dependence of LMR.** (a, b) Magnetic field dependent normalized resistance for the devices with channel lengths 150 nm and 850 nm. The corresponding FESEM images of the $Bi_2Se_3$ nanowire devices are shown in the insets. (c, d) The normalized resistances, normalized by the resistance at 12 K, vs. temperature curves extracted for CL 150 nm and 850nm as a function of B. Arrows indicate the points at which deviation in slopes started. Insets depict the isotherms of R(B)-R(3T) vs. B for 150 nm CL and R(B)-R(3.5T) vs. B for 850 nm CL.

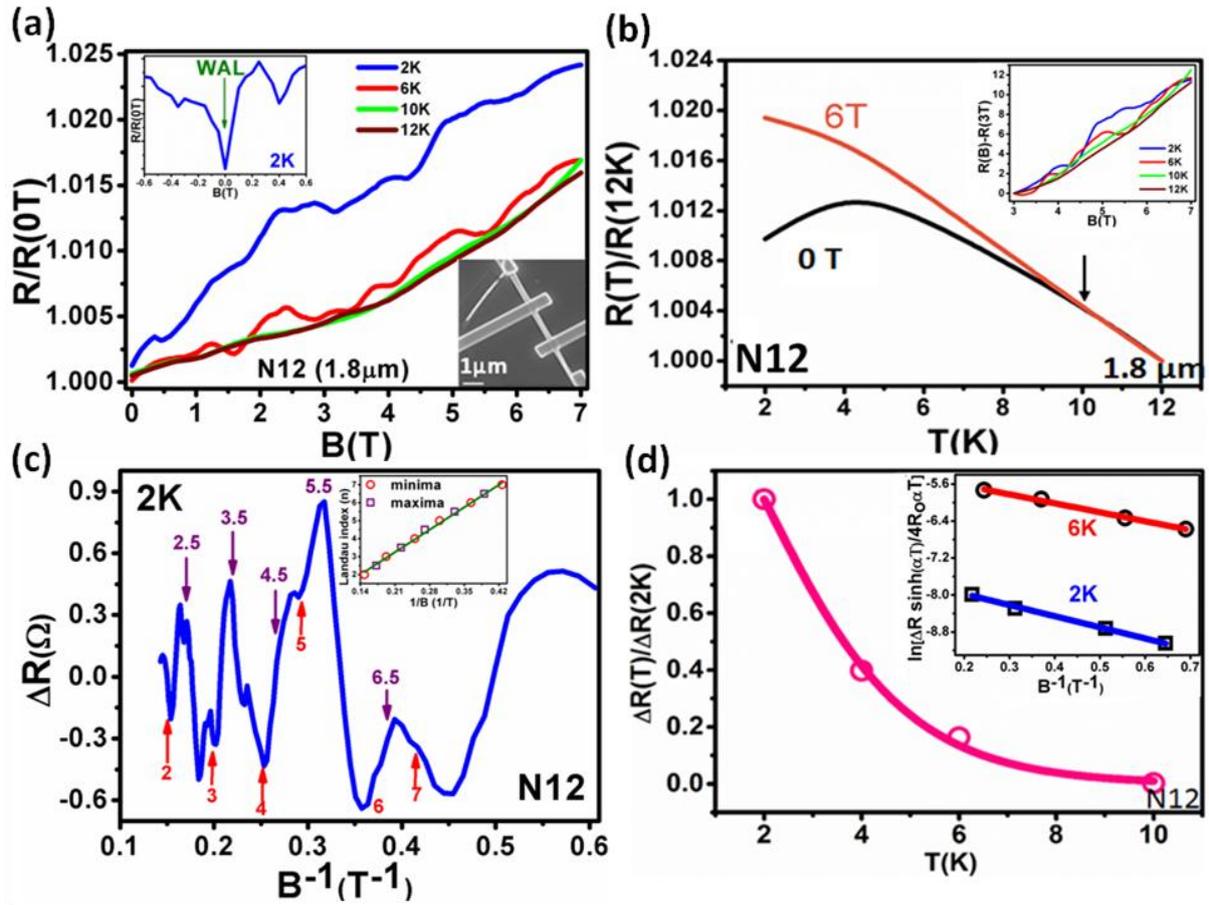

**Figure 3. Analysis of SdH oscillations in sample N12 (CL 1.8μm).** (a) Normalized MR showing WAL feature (inset) at 2K. Inset (down) shows the FESEM image of the device. (b) Normalized resistance vs. temperature curves extracted for CL 1.8μm as a function of B. Inset depict that isotherms of R(B)-R(3T) vs. B overlap for greater CL. (c) Variation in resistance (ΔR) as a function of inverse magnetic field. The arrows and numbers represent the values of maxima or minima used for plotting the Landau fan diagram as shown in the inset. (d) The temperature dependent plot of normalized resistance, normalized by the resistance measured at 2K, which represents scaled amplitude of SdH oscillations and a magnetic field 1.3 T was used to extract this data. The pink curve fit is used to estimate the cyclotron mass as explained in the text. Inset is the Dingle plot for 2K and 6K.

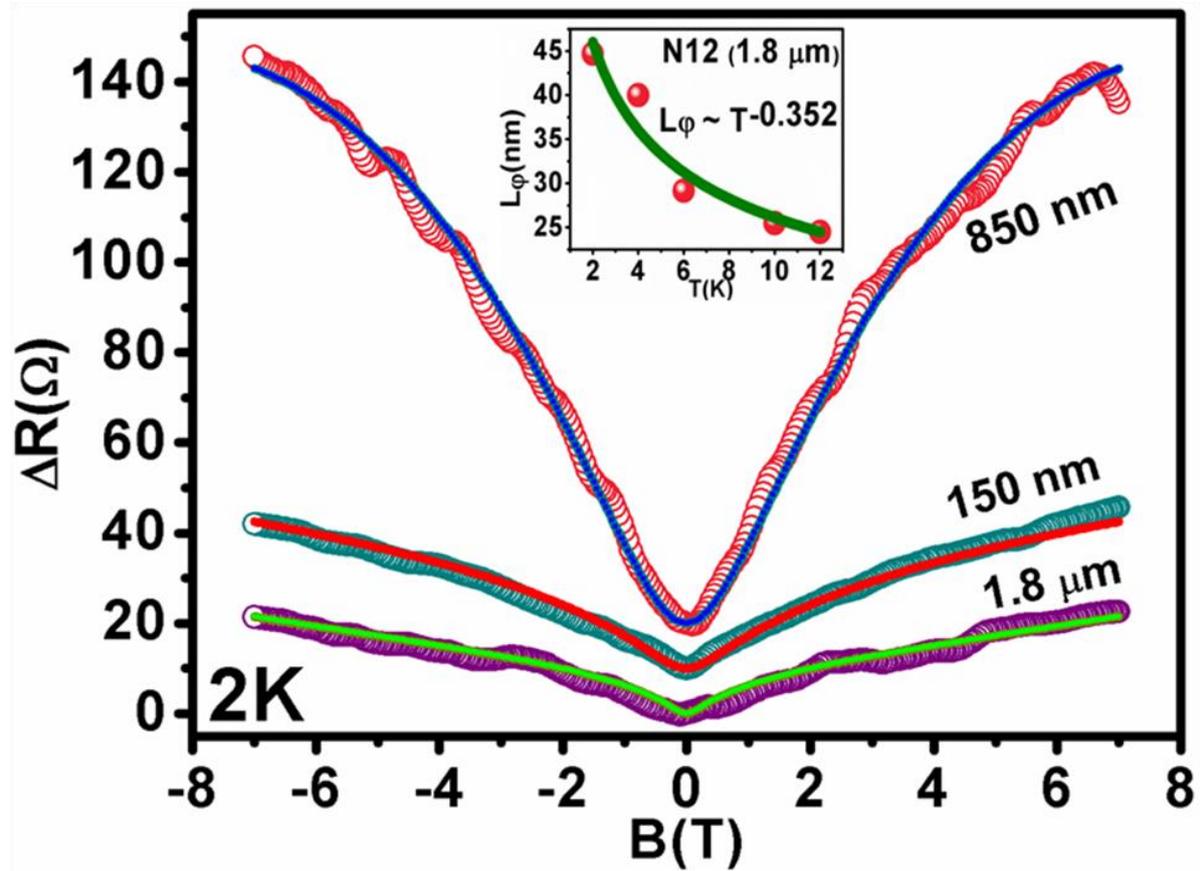

**Figure 4. Comparison of MR at 2K for the studied CLs.** The modified HLN was used to fit these curves. Inset is the phase coherence length decay plotted as a function of temperature.

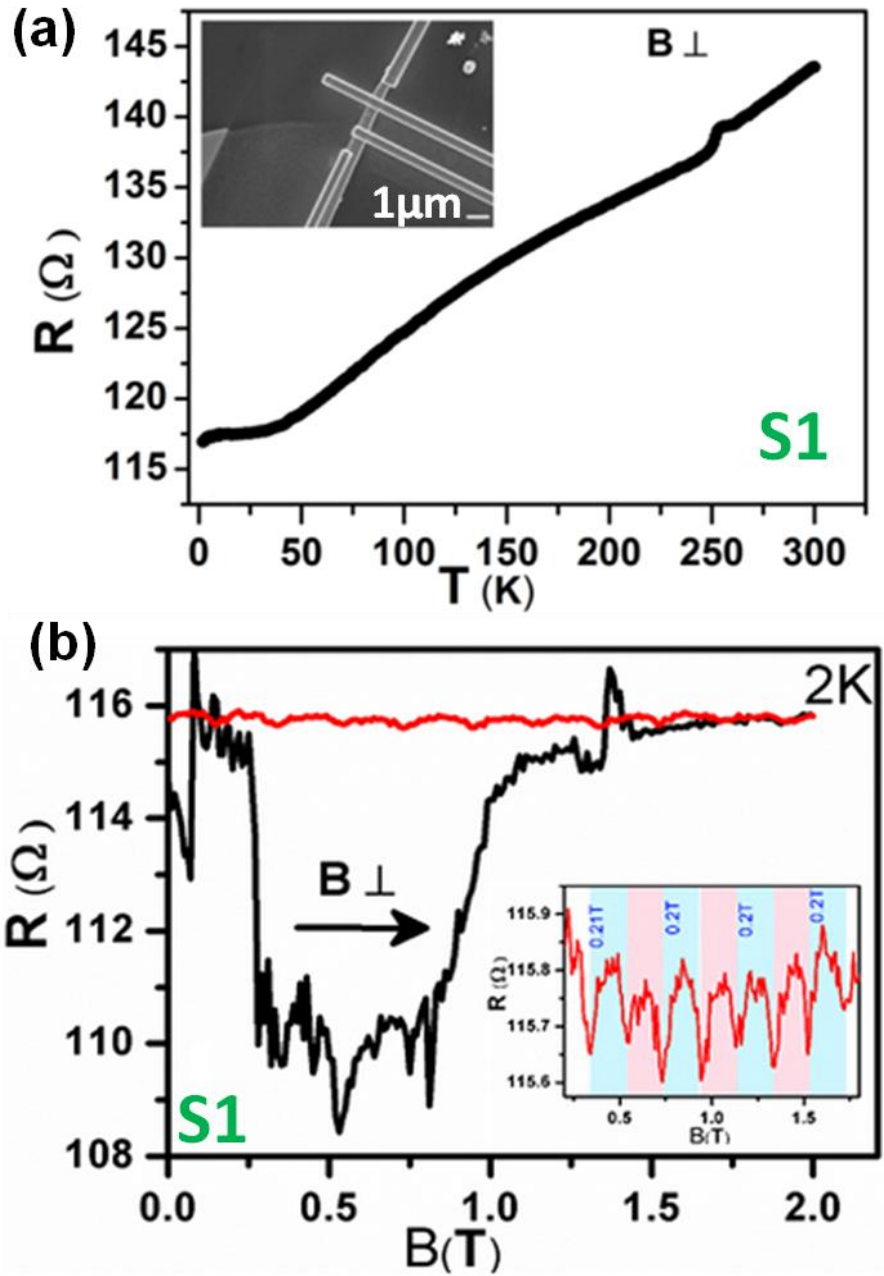

**Figure 5. Sample S1.** (a) RT curve shows metallic behavior, inset is device S1. (b) SdH oscillations in FIB fabricated device S1, a non periodic to periodic oscillations in presence of perpendicular magnetic field at the base temperature 2K. The arrow indicates sweep direction of B. Up sweep (0 to 2T, the black curve) and down sweep (2-0T, the red curve). Inset shows the periodic oscillations in MR observed for next up or down sweeps in B.

Supplementary Material

# FIB synthesis of Bi$_2$Se$_3$ 1D nanowires demonstrating the co-existence of Shubnikov-de Haas oscillations and linear magnetoresistance

**Biplab Bhattacharyya**[1,2], **Alka Sharma**[1,2], **V P S Awana**[1,2], **T. D. Senguttuvan**[1,2] and **Sudhir Husale**[1,2]*

[1]Academy of Scientific and Innovative Research (AcSIR), National Physical Laboratory, Council of Scientific and Industrial Research, Dr. K. S Krishnan Road, New Delhi-110012, India.

[2] National Physical Laboratory, Council of Scientific and Industrial Research, Dr. K. S Krishnan Road, New Delhi-110012, India.

*E-mail: husalesc@nplindia.org

**Contents:**

1. Table I: TSS comparison of the physical parameters.

2. Analysis for samples N10-11. Figures S1 and S2.

3. Smoothing filter used for experimental data. Figure S3.

4. Ramping rates of magnetic field.

# 1. Table I: TSS comparison of the physical parameters

| S. No. | $n_{2D}$ ($10^{10}$cm$^{-2}$) | $n_{3D}$ ($10^{15}$cm$^{-3}$) | $m_c$ ($m_o$) | $k_f$ (Å$^{-1}$) | $v_f$ ($10^5$ ms$^{-1}$) | $E_f$ (meV) | $T_D$ (K) | $\tau$ ($10^{-14}$ s) | $l$ (nm) | $\mu$ (cm$^2$V$^{-1}$s$^{-1}$) | $k_f l$ |
|---|---|---|---|---|---|---|---|---|---|---|---|
| 1. N10 | 57.6 | 665.4 | 0.023 | 0.027 | 13.7 | 243.6 | 5.93 | 20.5 | 280.85 | 15792 | 75.8 |
| N11 | 76.5 | 1007.17 | 0.026 | 0.031 | 13.89 | 282.4 | 8.86 | 13.7 | 190.293 | 9359.52 | 58.99 |
| N12 | 42.12 | 411.3 | 0.067 | 0.023 | 3.99 | 61.2 | 2.43 | 50.1 | 199.8 | 13034 | 46.5 |
| 2. | 9.6 | --- | 0.015 | 0.011 | 8.8 | 67 | 29 | 4.2 | 37 | 4866 | --- |
| 3. | 110 | --- | 0.119 | 0.038 | 3.67 | 92 | ----- | 31.1 | 114 | 4560 | --- |
| 4. | ---- | ---- | ---- | 0.03 | 4.2 | 78 | ---- | ---- | 219 | --- | 66 |
| 5. | 120 | --- | 0.16 | --- | 1.4 | --- | 1.65 | ---- | 105 | 8300 | --- |
| 6. | --- | --- | 0.11 | 0.031 | 4.2 | 90 (83) | --- | --- | --- | 1750 | 10 |
| 7. | 28.7 | --- | 0.06 | 0.019 | 3.6 | 45.7 | 16.9 | 7.2 | 26 | 2100 | 5 |
| 8. | --- | --- | 0.09 | --- | --- | --- | 6 | 40 | --- | 7800 | --- |
| 9. | 28 | 350 | 0.35 | 0.019 | 1.2 | --- | --- | 117 | 140 | 5840 | 50 |
| 10. (s1) | 35 | 310 | 0.09 | 0.021 | 2.7 | 37 | -- | --- | 31 | 2210 | 6.5 |
| (s2) | 32.8 | 283 | 0.09 | 0.0203 | 2.6 | 35 | | | 37 | 2800 | 7.5 |
| 11. | 39 | --- | 0.07 | 0.022 | 3.6 | 53 | --- | 130 | 460 | 13200 | --- |
| 12. (1) | 316 | --- | 0.18 | 0.063 | 3.96 | 176 | ---- | 40 | 161.7 | 3926 | --- |
| (2) | 178.6 | | 0.14 | 0.047 | 3.8 | 128 | | 24 | 92.6 | 3020 | --- |
| 13. | 110 | --- | 0.17 | 0.038 | -- | --- | 8.81 | 13.8 | --- | 8800 | --- |

**Reference (shown in the S. No.):**

## 2. Analysis for samples N10-11

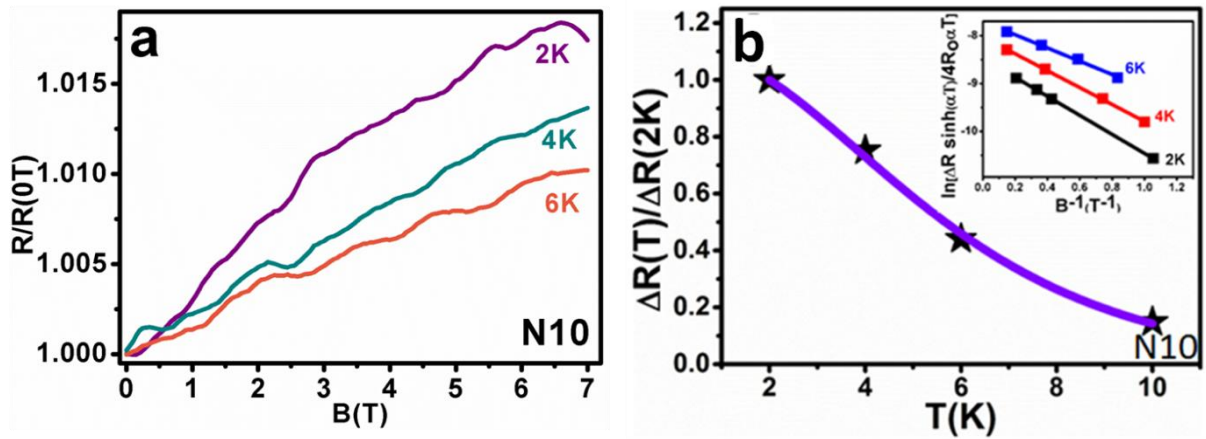

**Figure S1.** (a) The field dependent resistance oscillations, (b) temperature dependent normalized resistance with Dingle plots (inset) for device N10, note that B=0.8 T was used to plot the data.

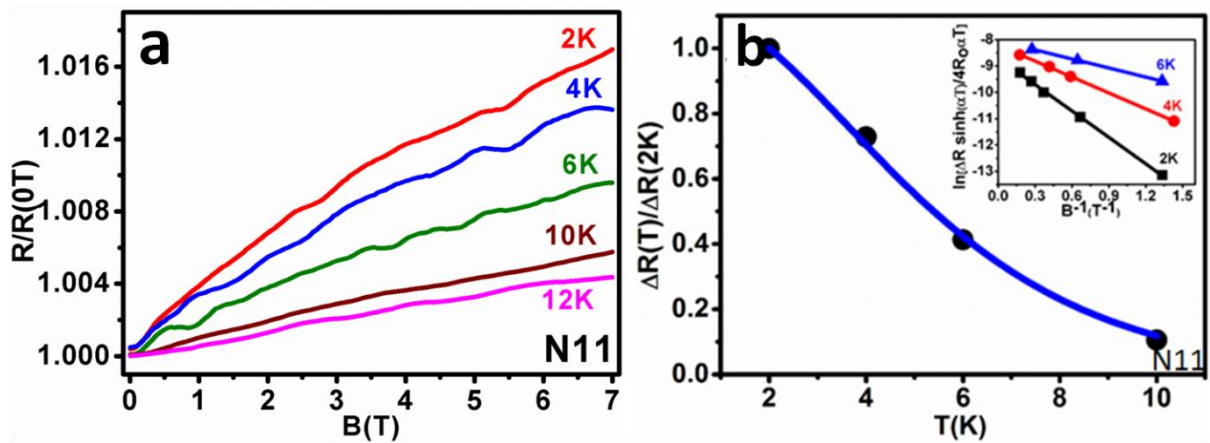

**Figure S2.** (a) The B field dependent MR measurements performed for device N11, (b) temperature dependent normalized resistance with Dingle plots (inset) where B = 0.85 T was used to plot the data.

## 3. Smoothing filter used for experimental data

The Savitzky-Golay digital filter in OriginPro 8 software is used to smooth the experimental data in order to increase the signal to noise ratio. A second order polynomial is used on a 10-point window to filter the data. A convolution process is used to fit successive data sets of 10 point each with a second degree polynomial using the method of linear least squares. The following figure shows the error bar in experimental and smoothed data, green markers are error bar and red circles are mean values of experimental and smoothed values.

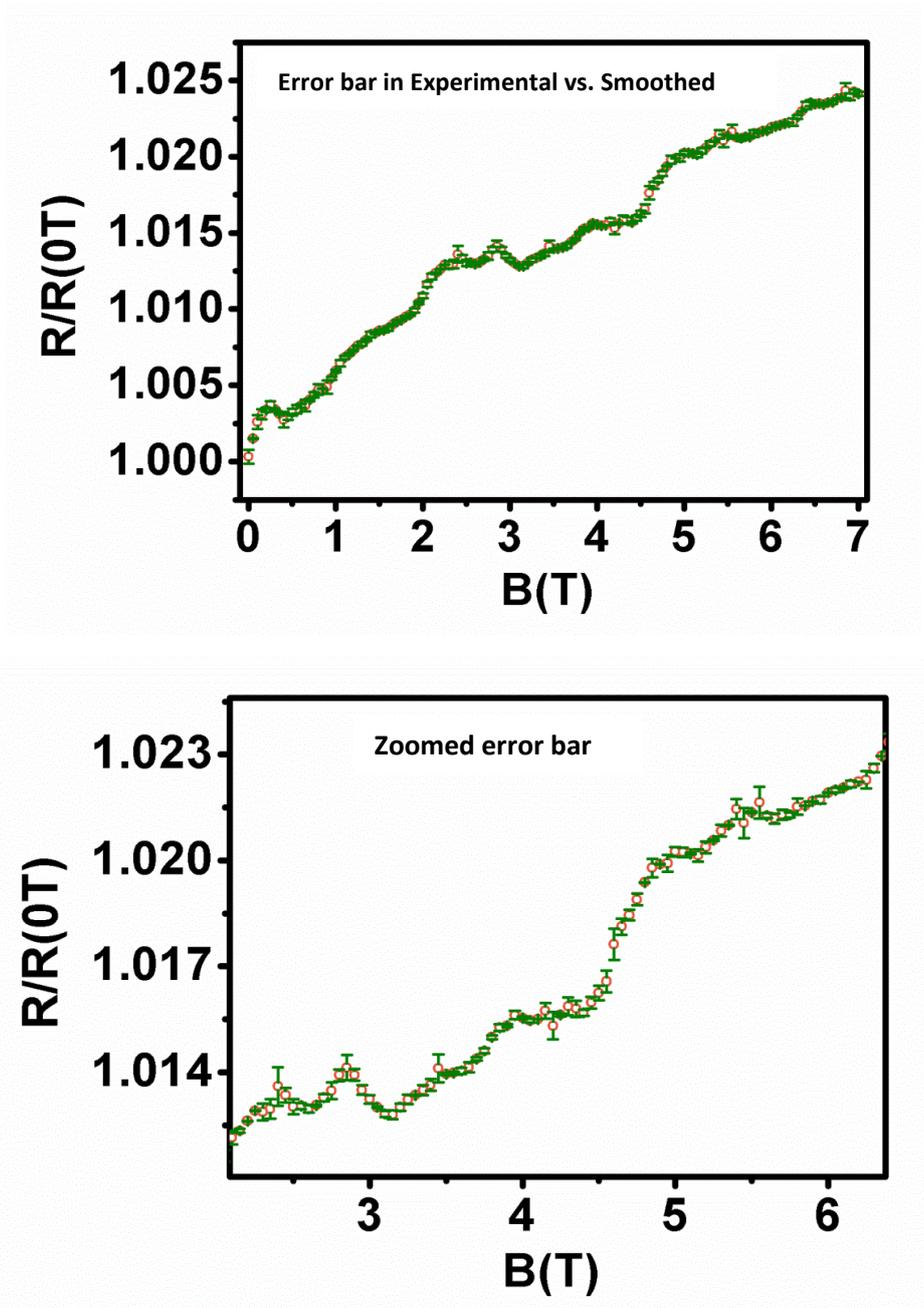

**Figure S3.** Error bar in experimental and smoothed data

## 4. Ramping rates of magnetic field

The Magnetic field ramping rates for devices N10-12 are as follows: 2K, 4K, 6K = 21.3 Oe/sec. and 10K, 12K = 35.1 Oe/sec respectively. For device S1: 0 to 2K: 0.87 Oe/sec. and for other sweeps it is around 5.5 Oe/sec.